\documentclass[conference]{IEEEtran}
\IEEEoverridecommandlockouts

\usepackage{amsmath,amssymb,amsthm,mathtools}
\usepackage{bm}
\usepackage{booktabs}
\usepackage{algorithm}
\usepackage{algorithmic}
\usepackage{graphicx}
\usepackage{cite}
\usepackage{hyperref}
\usepackage{color}

\newtheorem{proposition}{Proposition}

\newtheorem{remark}{Remark}

\DeclareMathOperator*{\argmin}{arg\,min}
\DeclareMathOperator{\KL}{KL}
\DeclareMathOperator{\MMD}{MMD}
\DeclareMathOperator{\ESS}{ESS}

\title{Sensor Design for Accuracy-Bounded Estimation\\via Maximum-Entropy Likelihood Synthesis}

\author{\IEEEauthorblockN{\bf Raktim Bhattacharya}
\IEEEauthorblockA{Department of Aerospace Engineering\\
Texas A\&M University\\
College Station, TX, USA\\
raktim@tamu.edu}}

\begin{document}
\maketitle

% ============================================================
\begin{abstract}
Determining the sensing architecture for large-scale spatio-temporal systems is nontrivial when estimation accuracy requirements are specified but sensor models are uncertain or unavailable. Classical design treats sensor placement and estimation sequentially, requiring valid forward models for each sensing modality. This paper inverts the design flow: given an estimation error budget, synthesize the measurement likelihood that enforces that budget while injecting minimal information beyond the dynamical prior. The likelihood is constructed via constrained optimization: among all posteriors satisfying a prescribed distributional accuracy bound relative to a desired target, select the one minimizing Kullback-Leibler divergence from the prior.
The solution is a maximum-entropy posterior in relative-entropy form, and the induced likelihood is recovered as the Radon-Nikodym derivative. The framework accommodates arbitrary discrepancy functionals and is instantiated here for Wasserstein distance, maximum mean discrepancy, $f$-divergences, moment constraints, and hybrid metrics. For each choice, we derive the discrete particle-level optimization problem, discuss the structure of the resulting convex or convex-relaxed program, and describe practical solvers with complexity scaling. A closed-form solution is given for the symmetric exponential-tilt case, and a distillation procedure converts nonparametric likelihood samples into compact parametric representations. A two-layer sensor design architecture embeds the synthesized likelihood in the predict-update loop of recursive state estimation, connecting accuracy budgets to physical sensor placement, precision, and configuration. Numerical experiments comparing four accuracy metrics on unimodal and multimodal scenarios confirm that the distributional accuracy constraints are reliably enforced and reveal how metric choice determines both the amount and spatial distribution of injected information.
\end{abstract}


% ============================================================
\section{Introduction}\label{sec:intro}

Bayesian filtering proceeds in two stages: prediction, which propagates the prior through a dynamical model, and update, which incorporates measurement information via a likelihood function. When the dynamical model is well understood but the measurement model is not, the update step becomes the weak link in the estimation pipeline. This situation arises in sensor fusion under model uncertainty \cite{BarShalom2001,Mahler2014}, in tracking with uncharacterized or heterogeneous sensors \cite{Gustafsson2010}, in data assimilation when observation operators are approximate \cite{Asch2016}, and in machine-learning-augmented state estimation where learned components lack calibrated likelihoods \cite{Gal2016}.

Large-scale spatio-temporal systems make the problem more difficult because the sensing architecture itself is a design variable. Environmental monitoring networks, distributed infrastructure systems, and multi-sensor arrays require decisions about sensor placement, measurement precision, and information fusion strategy. The classical workflow is sequential: derive a measurement likelihood from sensor physics, then optimize placement to minimize a predictive error criterion. This presupposes accurate forward models for each sensing modality. When such models are unavailable, unreliable, or prohibitively complex, or when system-level accuracy requirements must be satisfied without detailed component-level characterization, the forward design path fails. The approach developed here reverses the workflow. Given an estimation accuracy budget, we synthesize the measurement likelihood that enforces that budget while injecting minimal information beyond the dynamical prior. This establishes a direct path from error tolerance to sensing architecture.

The standard response to likelihood uncertainty is robustification: minimax filters, covariance inflation, or adaptive methods that tune likelihood parameters online \cite{Mehra1970,Mehra1972,Sarkka2023}. These approaches presuppose a parametric likelihood family and seek protective settings within it. An alternative philosophy, pursued here, is to \emph{synthesize} a likelihood from posterior design goals rather than from a forward sensor model.

The starting observation is that in many applications a practitioner can articulate a meaningful prior $\pi_k^-$ (from the prediction step) and a desired posterior $\pi_k^\star$ (from external information, physics-based constraints, or a reference estimator), but cannot reliably specify the likelihood function that should connect them. Traditional Bayesian filtering derives the posterior from a likelihood via Bayes' rule. This paper inverts the process: solve an optimization problem to find a posterior $\pi_k^+$ that is close to the desired target $\pi_k^\star$ while remaining conservative relative to the prior $\pi_k^-$, then recover the implied likelihood as a Radon-Nikodym derivative. The likelihood becomes a design artifact, synthesized to achieve distributional goals, rather than a physical sensor model derived from first principles. The synthesized likelihood, once parameterized, specifies what information structure must be provided by the sensing architecture and can be inverted to determine sensor placement, precision, and configuration that realize the desired estimation performance.

\begin{remark}[Data-free construction]
The entire framework operates without measurement data. The likelihood is not estimated from observations; it is synthesized from an accuracy specification and the prior. No sensor hardware, measurement noise model, or calibration data is required. The output is a prescription for what a sensor \emph{should} measure, not a model of what it \emph{does} measure.
\end{remark}

\subsection{Related Work}

Maximum-entropy distribution construction under constraints originates with Jaynes \cite{Jaynes1957a,Jaynes1957b} and Csisz\'{a}r \cite{Csiszar1975}; its connection to exponential families is classical \cite{Wainwright2008}.
Distributionally robust filtering using relative entropy constraints was studied by Levy and Nikoukhah \cite{Levy2013}, and Wasserstein distributionally robust optimization by Shafieezadeh-Abadeh et al.\ \cite{Shafieezadeh2019}.
Optimal transport enters filtering through ensemble transforms \cite{Reich2013,Reich2015}, where transport maps move particles from prior to posterior; our formulation uses transport distance as a \emph{constraint} on accuracy rather than as the update mechanism.
Particle filtering \cite{Doucet2001,Chopin2020} provides the computational backbone, and the weight-rebalancing view connects to approximate Bayesian computation \cite{Marin2012}, though our approach is optimization-based.

\subsection{Contributions}
The main contributions are:
\begin{enumerate}
\item A discrepancy-agnostic framework for likelihood synthesis via KL-minimization subject to distributional accuracy constraints, with particle-level formulations for nonparametric updates and likelihood recovery (Section~\ref{sec:framework}).
\item Instantiation for Wasserstein distance, MMD, $f$-divergences, moment constraints, and hybrid metrics, with structural analysis and complexity scaling (Section~\ref{sec:metrics}).
\item A closed-form exponential-tilt solution for the symmetric case and a distillation procedure for parametric likelihood recovery (Sections~\ref{sec:symmetric},~\ref{sec:sensor_design}).
\item A two-layer sensor design architecture connecting accuracy-bounded estimation to physical sensor configuration (Section~\ref{sec:sensor_design}).
\item Numerical experiments comparing four accuracy metrics on unimodal and multimodal scenarios (Section~\ref{sec:experiments}).
\end{enumerate}

% ============================================================
\section{General Maximum-Entropy Framework}\label{sec:framework}

\subsection{Problem Statement}

Let $(\mathcal{X}, d)$ be a Polish state space and $\mathcal{P}(\mathcal{X})$ the set of Borel probability measures on $\mathcal{X}$.
At time step $k$, let $\pi_k^- \in \mathcal{P}(\mathcal{X})$ be the prior (output of the prediction step) and $\pi_k^\star \in \mathcal{P}(\mathcal{X})$ a desired posterior encoding a tracking accuracy specification.
In traditional Bayesian filtering, one measures data $y_k$, derives a likelihood $L_k(x) = p(y_k|x)$ from a sensor model, and applies Bayes' rule. Here we invert the workflow: the desired posterior is specified, and the likelihood is synthesized. No measurement data is involved; the likelihood is a design variable.

A likelihood function $L_k : \mathcal{X} \to \mathbb{R}_+$ induces a posterior in Bayesian form:
\begin{equation}
\pi_k^+(dx) = \frac{L_k(x)\, \pi_k^-(dx)}{\int L_k(\xi)\, \pi_k^-(d\xi)}.
\label{eq:bayes_update}
\end{equation}

Given $\pi_k^-$ and target $\pi_k^\star$, the likelihood synthesis problem finds a posterior close to $\pi_k^\star$ while injecting minimal information beyond the prior.
For a discrepancy functional $\mathcal{D} : \mathcal{P}(\mathcal{X}) \times \mathcal{P}(\mathcal{X}) \to [0, \infty]$, the maximum-entropy likelihood synthesis problem is:
\begin{equation}
\boxed{
\min_{\pi \ll \pi_k^-}\; D_{\KL}(\pi \| \pi_k^-) \quad \text{s.t.} \quad \mathcal{D}(\pi, \pi_k^\star) \le \varepsilon_k.
}
\label{eq:generic_problem}
\end{equation}

The scalar $\varepsilon_k > 0$ is the accuracy budget.
The constraint set $\mathcal{C}_{\varepsilon_k} = \{\pi \in \mathcal{P}(\mathcal{X}) : \mathcal{D}(\pi, \pi_k^\star) \le \varepsilon_k\}$ is convex for Wasserstein, MMD, and moment constraints (ensuring a unique solution by strict convexity of the KL objective), but not generally for $f$-divergences.

\subsection{Likelihood Recovery}

\begin{proposition}[Likelihood from the optimal posterior]\label{prop:likelihood_recovery}
Suppose $\pi_k^{\mathrm{opt}}$ solves \eqref{eq:generic_problem} and $\pi_k^{\mathrm{opt}} \ll \pi_k^-$.
Then the likelihood function
\begin{equation}
L_k^{\mathrm{opt}}(x) \propto \frac{d\pi_k^{\mathrm{opt}}}{d\pi_k^-}(x)
\label{eq:rn_ratio}
\end{equation}
is the unique (up to positive scaling) nonnegative function such that the update in Bayesian form \eqref{eq:bayes_update} with $L_k = L_k^{\mathrm{opt}}$ yields $\pi_k^+ = \pi_k^{\mathrm{opt}}$.
\end{proposition}

The proof follows directly from \eqref{eq:bayes_update}: setting $\pi_k^+ = \pi_k^{\mathrm{opt}}$ and rearranging gives $L_k(x) \propto (d\pi_k^{\mathrm{opt}} / d\pi_k^-)(x)$.

\subsection{Particle Instantiation (Generic Discrepancy)}\label{sec:particle_generic}

For computational implementation, we specialize to discrete distributions.
In sequential Monte Carlo, prior particles propagate forward through dynamics, and measurement updates correspond to reweighting.
Classically, weights are updated via Bayes' rule with a given likelihood.
Here, we solve the constrained optimization to synthesize the weights directly, then recover the implied likelihood.

Represent the prior and desired posterior as weighted particle sets:
\begin{equation}
\pi_k^- = \sum_{i=1}^{N} w_{k,i}^-\, \delta_{x_{k,i}}, \qquad
\pi_k^\star = \sum_{j=1}^{M} v_{k,j}\, \delta_{z_{k,j}},
\end{equation}
where $w_k^- \in \Delta_N = \{ w \in \mathbb{R}_+^N : \sum_i w_i = 1 \}$ and $v_k \in \Delta_M$.

The particle-level synthesis problem is:
\begin{equation}
\min_{w_k \in \Delta_N}\; \sum_{i=1}^{N} w_{k,i} \log \frac{w_{k,i}}{w_{k,i}^-}
\quad \text{s.t.} \quad \mathcal{D}_N(w_k, v_k) \le \varepsilon_k,
\label{eq:generic_particle_problem}
\end{equation}
where $\mathcal{D}_N$ is the discrete counterpart of $\mathcal{D}$.

The likelihood recovered on the particle support is:
\begin{equation}
L_{k,i}^{\star} \propto \frac{w_{k,i}^{\mathrm{opt}}}{w_{k,i}^-}, \qquad i = 1, \ldots, N.
\label{eq:particle_likelihood}
\end{equation}

% ============================================================
\section{Specific Accuracy Metrics}\label{sec:metrics}

We now instantiate \eqref{eq:generic_problem} with concrete choices for the discrepancy $\mathcal{D}$.
Different metrics induce different optimization structures; the choice depends on the operational meaning of accuracy in the application.

\subsection{Wasserstein Distance}\label{sec:wasserstein}

For $p \ge 1$, the $p$-Wasserstein distance between $\mu, \nu \in \mathcal{P}(\mathcal{X})$ is:
\begin{equation}
W_p(\mu, \nu) = \left( \inf_{\gamma \in \Pi(\mu, \nu)} \int d(x,y)^p\, d\gamma(x,y) \right)^{1/p},
\end{equation}
where $\Pi(\mu, \nu)$ is the set of couplings with marginals $\mu$ and $\nu$.

Setting $\mathcal{D}(\pi, \pi_k^\star) = W_p(\pi, \pi_k^\star)$ and passing to the particle representation, the coupling becomes a matrix $\Gamma_k \in \mathbb{R}_+^{N \times M}$ and the problem is:
\begin{equation}
\begin{aligned}
\min_{w_k, \Gamma_k}\; & \sum_{i=1}^{N} w_{k,i} \log \frac{w_{k,i}}{w_{k,i}^-} \\
\text{s.t.}\; & \Gamma_k \mathbf{1}_M = w_k, \quad \Gamma_k^\top \mathbf{1}_N = v_k, \\
& \langle C_k, \Gamma_k \rangle \le \varepsilon_k^p, \\
& w_k \in \Delta_N, \quad \Gamma_k \ge 0,
\end{aligned}
\label{eq:wasserstein_program}
\end{equation}
where $C_{k,ij} = d(x_{k,i}, z_{k,j})^p$ is the ground cost matrix.

\begin{remark}[Structure]
Problem \eqref{eq:wasserstein_program} is jointly convex in $(w_k, \Gamma_k)$.
The marginal constraints enforce that $\Gamma_k$ transports mass from $w_k$ to $v_k$; the first marginal $\Gamma_k \mathbf{1}_M = w_k$ links the transport plan to the posterior.
The Wasserstein constraint $\langle C_k, \Gamma_k \rangle \le \varepsilon_k^p$ bounds the transport cost, ensuring the posterior is geometrically close to the target.
\end{remark}

\paragraph*{Sinkhorn solver.}
The Sinkhorn algorithm \cite{Cuturi2013,Peyre2019} approximates $W_p$ via entropic regularization at $\mathcal{O}(NM)$ per iteration.
The overall problem \eqref{eq:wasserstein_program} is solved via an outer bisection loop on the Lagrange multiplier $\lambda_k$ associated with the Wasserstein constraint.
At each candidate $\lambda_k$, the inner problem is an entropy-regularized transport computation solved by Sinkhorn iterations; the marginal constraint $\Gamma_k \mathbf{1}_M = w_k$ links the optimal transport plan to the posterior weights.
Bisection terminates when $|W_p(w_k(\lambda_k), v_k) - \varepsilon_k|$ falls below a prescribed tolerance, typically requiring $\mathcal{O}(\log(1/\delta))$ outer iterations for tolerance $\delta$.

Wasserstein distance measures the minimum cost of physically transporting probability mass from the posterior to the target in state space.
It is the natural accuracy metric when state-space displacement is the operationally relevant notion of error, for example in spatial localization or trajectory estimation.
It remains well-defined even when the prior and target have non-overlapping supports, where density-ratio-based divergences become infinite.

\subsection{Maximum Mean Discrepancy (MMD)}\label{sec:mmd}

Let $\kappa : \mathcal{X} \times \mathcal{X} \to \mathbb{R}$ be a positive-definite kernel with RKHS $\mathcal{H}_\kappa$.
The MMD between $\mu$ and $\nu$ is $\MMD(\mu, \nu) = \| m_\mu - m_\nu \|_{\mathcal{H}_\kappa}$, where $m_\mu = \int \kappa(\cdot, x)\, d\mu(x)$ is the kernel mean embedding.
For discrete distributions, the squared MMD expands to:
\begin{equation}
\MMD^2(w_k, v_k) = w_k^\top K_{xx} w_k - 2 w_k^\top K_{xz} v_k + v_k^\top K_{zz} v_k,
\label{eq:mmd_discrete}
\end{equation}
where $[K_{xx}]_{ij} = \kappa(x_{k,i}, x_{k,j})$, $[K_{xz}]_{ij} = \kappa(x_{k,i}, z_{k,j})$, and $[K_{zz}]_{ij} = \kappa(z_{k,i}, z_{k,j})$.

The synthesis problem becomes:
\begin{equation}
\begin{aligned}
\min_{w_k \in \Delta_N}\; & \sum_{i=1}^{N} w_{k,i} \log \frac{w_{k,i}}{w_{k,i}^-} \\
\text{s.t.}\; & w_k^\top K_{xx} w_k - 2 w_k^\top K_{xz} v_k \le \varepsilon_k^2 - v_k^\top K_{zz} v_k.
\end{aligned}
\label{eq:mmd_program}
\end{equation}
The constraint is rewritten with the constant $v_k^\top K_{zz} v_k$ on the right-hand side.
Since $K_{xx} \succeq 0$, the constraint is convex, and \eqref{eq:mmd_program} is a convex program with a unique solution.
MMD compares distributions through their kernel mean embeddings, providing a smooth, differentiable discrepancy that avoids explicit transport plans.
The performance depends on kernel choice: a Gaussian kernel with bandwidth $\sigma$ is sensitive to differences at scale $\sigma$.

\subsection{\texorpdfstring{$f$}{f}-Divergence Metrics}\label{sec:fdiv}

For a convex function $f : \mathbb{R}_+ \to \mathbb{R}$ with $f(1) = 0$, the $f$-divergence of $\mu$ with respect to $\nu$ is:
\begin{equation}
D_f(\mu \| \nu) = \int f\!\left(\frac{d\mu}{d\nu}\right) d\nu.
\end{equation}
Common instances include KL ($f(t) = t \log t$), reverse KL ($f(t) = -\log t$), $\chi^2$ ($f(t) = (t-1)^2$), and Jensen-Shannon divergence.
Using $\mathcal{D}(\pi, \pi_k^\star) = D_f(\pi \| \pi_k^\star)$ requires $\pi \ll \pi_k^\star$, i.e., the posterior must be absolutely continuous with respect to the target.
For particle filters, this means prior and target must share support points.
\begin{equation}
D_f(w_k \| v_k) = \sum_{j} v_{k,j}\, f\!\left(\frac{w_{k,j}}{v_{k,j}}\right),
\end{equation}
where the sum runs over shared support points and $w_{k,j}$ is the posterior weight at the $j$-th shared location.
If supports do not overlap, $D_f$ is typically infinite or undefined.
In our particle implementation, the $\chi^2$-divergence constraint is evaluated by interpolating the target weights onto the prior support via kernel smoothing, so that every prior particle has a well-defined target weight; this introduces a small approximation but avoids the degenerate case.
This makes $f$-divergences best suited for shared-support settings, such as hypothesis testing, where the density ratio has direct operational meaning as a likelihood ratio.

\subsection{Moment and Feature Constraints}\label{sec:moments}

A lightweight choice is to constrain moments or features:
\begin{equation}
\mathcal{D}(\pi, \pi_k^\star) := \|\mathbb{E}_\pi[\phi(x)] - \mathbb{E}_{\pi_k^\star}[\phi(x)]\|,
\end{equation}
for a feature map $\phi : \mathcal{X} \to \mathbb{R}^d$ and norm $\|\cdot\|$.
Choosing $\phi(x) = x$ constrains the mean; $\phi(x) = (x, xx^\top)$ constrains both mean and covariance; nonlinear $\phi$ constrains application-specific functionals.

The discrete constraint is affine in $w_k$ inside the norm:
\begin{equation}
\left\| \sum_{i=1}^N w_{k,i}\, \phi(x_{k,i}) - \sum_{j=1}^M v_{k,j}\, \phi(z_{k,j}) \right\| \le \varepsilon_k.
\end{equation}
For the Euclidean norm this is a second-order cone constraint; for $\|\cdot\|_1$ or $\|\cdot\|_\infty$, it can be reformulated as linear inequalities via auxiliary variables.
The resulting program is computationally efficient but does not guarantee full distributional fidelity: the maximum-entropy principle selects among all distributions satisfying the moment bounds, which may differ from the target in shape or tail behavior.

\subsection{Hybrid Metrics}\label{sec:hybrid}

The framework supports multiple constraints simultaneously:
\begin{equation}
\min_{\pi \ll \pi_k^-}\; D_{\KL}(\pi \| \pi_k^-) \quad \text{s.t.} \quad \mathcal{D}_q(\pi, \pi_k^\star) \le \varepsilon_{k,q}, \quad q = 1, \ldots, Q.
\label{eq:multi_constraint}
\end{equation}
Alternatively, discrepancies combine via $\mathcal{D} = \sum_q \alpha_q \mathcal{D}_q$ with a single budget $\varepsilon_k$.

% ============================================================
\section{Special Symmetric Case}\label{sec:symmetric}

When the desired posterior is a Dirac at the origin and the accuracy metric is a second-moment constraint $\mathbb{E}_{\pi_e}[\|e\|^2] \le \varepsilon_k^2$ (with $e = x - x^\star$), the solution is an exponential tilt.

\begin{proposition}[Exponential tilt]\label{prop:exp_tilt}
Under the second-moment constraint, the solution to \eqref{eq:generic_problem} is:
\begin{equation}
\frac{d\pi_{e,k}^{\mathrm{opt}}}{d\pi_{e,k}^-}(e)
= \frac{\exp(-\lambda_k \|e\|^2)}{Z_k(\lambda_k)},
\label{eq:symmetric_tilt}
\end{equation}
where $\lambda_k \ge 0$ is the Lagrange multiplier and $Z_k(\lambda_k) = \int \exp(-\lambda_k \|e\|^2)\, d\pi_{e,k}^-$.
\end{proposition}

This is a standard result; the same structure appears in the robust filtering framework of Levy and Nikoukhah \cite{Levy2013}.
The proof follows from the KKT conditions of \eqref{eq:generic_problem} with the quadratic constraint.
For particles:
\begin{equation}
w_{k,i}^+(\lambda) = \frac{w_{k,i}^-\, \exp(-\lambda \|e_{k,i}\|^2)}{\sum_{\ell=1}^N w_{k,\ell}^-\, \exp(-\lambda \|e_{k,\ell}\|^2)}.
\label{eq:closed_form_weights}
\end{equation}

The multiplier $\lambda_k$ is found by scalar root-finding on $g(\lambda) = \sum_i w_{k,i}^+(\lambda) \|e_{k,i}\|^2 - \varepsilon_k^2 = 0$.
This costs $\mathcal{O}(I_\lambda N)$ per update but produces a radially symmetric likelihood, which may not match asymmetric measurement geometries.
When the prior is Gaussian, $\pi_{e,k}^- = \mathcal{N}(0, P_k^-)$, the optimal posterior is $\mathcal{N}(0, ((P_k^{-})^{-1} + 2\lambda_k I)^{-1})$, recovering a Kalman-like update.

% ============================================================
\section{Likelihood Parameterization and Sensor Design}\label{sec:sensor_design}

The maximum-entropy framework developed above synthesizes an ideal likelihood $L_k^\star$ from a performance specification.
This section first describes how to distill the nonparametric likelihood into a parametric family, then shows how the parameterized likelihood serves as the interface between estimation requirements and physical sensor design.

\subsection{Parametric Distillation}\label{sec:parameterization}

Equation \eqref{eq:particle_likelihood} provides nonparametric likelihood values on the particle support.
When a compact or interpretable likelihood is needed, distillation into a parametric family is appropriate.
Two natural choices are Gaussian mixtures,
\begin{equation}
L_k(x; \theta_k) = \sum_{m=1}^{M_g} \alpha_{k,m}\, \mathcal{N}(x; \mu_{k,m}, \Sigma_{k,m}),
\label{eq:mog_likelihood}
\end{equation}
and exponential families $L_k(x; \eta_k) \propto \exp(\eta_k^\top \phi(x))$, of which \eqref{eq:symmetric_tilt} is a special case.
Parameters are fit via weighted least squares in log domain:
\begin{equation}
\min_{\theta_k}\; \sum_{i=1}^{N} \bar{w}_{k,i} \left( \log L_{k,i}^\star - \log L_k(x_{k,i}; \theta_k) \right)^2,
\label{eq:distillation}
\end{equation}
where $\bar{w}_{k,i} = w_{k,i}^{\mathrm{opt}} / \sum_\ell w_{k,\ell}^{\mathrm{opt}}$ are the normalized optimal weights and $\theta_k$ collects all parameters of the chosen family (e.g., component means, covariances, and weights for a Gaussian mixture).
Distillation is optional; the nonparametric likelihood \eqref{eq:particle_likelihood} can be used directly for weight updates.
Distillation is useful when the likelihood must be communicated to another estimator, stored compactly, evaluated off the particle support, or when physical interpretability of sensor parameters is desired.

\subsection{Desired Posterior as a Performance Specification}\label{sec:desired_posterior}

The desired posterior $\pi_k^\star$ is a performance specification, not privileged information about the true state.
Given prior particles $\{x_{k,i}, w_{k,i}^-\}_{i=1}^N$, let $\hat{x}_k$ be any available reference (the prior mean, an EKF estimate, or a fusion input).
The simplest specification is $\pi_k^\star = \delta_{\hat{x}_k}$, under which the $W_2$ constraint becomes:
\begin{equation}
W_2(\pi_k^+, \delta_{\hat{x}_k}) = \sqrt{\sum_{i=1}^N w_{k,i}^+\, \|x_{k,i} - \hat{x}_k\|^2} \le \varepsilon_k,
\label{eq:rms_bound}
\end{equation}
a direct RMS error bound.
More generally, $\pi_k^\star = \mathcal{N}(\hat{x}_k, \bar{P}_k)$ specifies a covariance budget. No knowledge of the true state is needed.
Other choices include a fusion input from a partner node (enforcing track agreement) or a reference filter output (for design-time distillation).
The framework assumes that $\pi_k^\star$ is realizable, i.e., the constraint $\mathcal{D}(\pi, \pi_k^\star) \le \varepsilon_k$ is feasible for some $\pi \ll \pi_k^-$. Systematic methods for constructing and validating $\pi_k^\star$ from operational requirements remain an open problem.

\subsection{Two-Layer Architecture}\label{sec:two_layer}
The connection between accuracy-bounded estimation and sensor design is a two-layer decomposition.

\paragraph{Layer 1: Information design.}
Solve \eqref{eq:generic_problem} to obtain $\pi_k^{\mathrm{opt}}$ and the target likelihood $L_{k,i}^\star = w_{k,i}^{\mathrm{opt}} / w_{k,i}^-$.
This layer determines what the sensor must achieve, expressed as a likelihood shape over state space.

\paragraph{Layer 2: Sensor configuration.}
Choose physical sensor parameters $\theta_k \in \Theta$ (positions, bandwidths, beam patterns, etc.) so that the realizable likelihood $L_k(x; \theta_k)$ approximates $L_k^\star$.
When $L_k(\cdot\,; \theta_k) = L_k^\star(\cdot)$ exactly, the accuracy constraint is met by construction.
In practice, a realizability gap exists:
\begin{equation}
\mathcal{D}(\pi_k^+, \pi_k^\star) \le \varepsilon_k + \Delta_k(\theta_k),
\label{eq:realizability_gap}
\end{equation}
where $\Delta_k(\theta_k) \ge 0$ depends on how well the sensor likelihood approximates the ideal one.
Layer~2 minimizes this gap.

\subsection{Sensor Design for Recursive State Estimation}\label{sec:placement}

The preceding subsections define the two layers abstractly.
We now describe the concrete sensor parameterization, the fitting procedure, and how both layers integrate into a recursive predict-update-resample loop with optimal sensing.

\paragraph{Sensor likelihood model.}
Suppose the sensor suite consists of $R$ sensors, each producing a Gaussian kernel likelihood centered at its position $s_{k,r} \in \mathcal{X}$ with bandwidth $h_{k,r}$.
Range sensors, radar, sonar, and thermal imagers all produce approximately Gaussian likelihoods centered at the measurement source, so a mixture of such kernels is a natural multi-sensor family:
\begin{equation}
L_k(x; \theta_k) = \sum_{r=1}^R \alpha_{k,r}\, \exp\!\left(-\frac{\|x - s_{k,r}\|^2}{2 h_{k,r}^2}\right),
\label{eq:sensor_mixture}
\end{equation}
with sensor parameters $\theta_k = (\bm{s}_k, \bm{\alpha}_k, \bm{h}_k)$ comprising positions $\bm{s}_k = (s_{k,1}, \ldots, s_{k,R})$, mixing weights $\bm{\alpha}_k \in \Delta_R$, and per-component bandwidths $\bm{h}_k \in \mathbb{R}_+^R$.

\paragraph{Scale-invariant fitting.}
Since the likelihood is defined only up to a positive scaling (the normalizing constant in Bayes' rule absorbs any multiplicative factor), the fitting objective must be scale-invariant.
Define the log-domain residual $r_i(\theta_k) = \log L_{k,i}^\star - \log L_k(x_{k,i}; \theta_k)$ and its weighted mean $\bar{c}(\theta_k) = \sum_{i} \bar{w}_{k,i}\, r_i(\theta_k)$, which absorbs the unknown log-scale offset.
The sensor design problem minimizes the scale-invariant fitting error:
\begin{equation}
\min_{\theta_k \in \Theta}\; \sum_{i=1}^N \bar{w}_{k,i}\!\left(r_i(\theta_k) - \bar{c}(\theta_k)\right)^2
+ \mathcal{J}(\theta_k, \theta_{k-1}),
\label{eq:sensor_design}
\end{equation}
which specializes the generic distillation objective \eqref{eq:distillation} to the sensor mixture family \eqref{eq:sensor_mixture}, with the centered residual $r_i - \bar{c}$ replacing the raw log-difference to absorb the arbitrary normalization.
The first term measures how well the parametric likelihood matches the \emph{shape} of $L_k^\star$ and $\mathcal{J}$ encodes operational costs (sensor relocation, coverage overlap).
The fitting weights $\bar{w}_{k,i} = w_{k,i}^{\mathrm{opt}} / \sum_\ell w_{k,\ell}^{\mathrm{opt}}$ concentrate the approximation effort where the optimal posterior places mass.
The operational cost $\mathcal{J}$ is application-specific (e.g., sensor relocation penalties, coverage costs); in the experiments below we set $\mathcal{J} = 0$ to isolate the likelihood fitting quality.
To handle the simplex constraint on $\bm{\alpha}_k$ and positivity of $\bm{h}_k$, we reparameterize via unconstrained variables: $\alpha_{k,r} = e^{\ell_{k,r}} / \sum_{r'} e^{\ell_{k,r'}}$ (softmax) and $h_{k,r} = e^{\eta_{k,r}}$ (log-bandwidth), yielding an unconstrained optimization over $(\bm{s}_k, \bm{\ell}_k, \bm{\eta}_k) \in \mathbb{R}^{(d+2)R}$.
Multi-start gradient-based optimization (L-BFGS-B) with initializations from quantile placement and Silverman bandwidth provides reliable convergence.
Algorithm~\ref{alg:layer2} summarizes the procedure.

\begin{algorithm}[!ht]
\caption{Layer~2: Scale-Invariant Sensor Fitting}
\label{alg:layer2}
\begin{algorithmic}[1]
\REQUIRE Prior particles $\{x_{k,i}, w_{k,i}^-\}_{i=1}^N$, optimal weights $w_{k,i}^{\mathrm{opt}}$ from Layer~1, number of sensors~$R$, restarts~$T$.
\STATE Ideal likelihood: $L_{k,i}^\star \leftarrow w_{k,i}^{\mathrm{opt}} / w_{k,i}^-$.
\STATE Initialize $\bm{s}^{(0)}$ at quantiles of $(x_{k,i}, \bar{w}_{k,i})$; set $h_r^{(0)}$ via Silverman's rule; set $\bm{\ell}^{(0)} = \bm{0}$.
\FOR{$t = 1, \ldots, T$}
\STATE If $t > 1$: perturb $\bm{s}^{(0)} \!+\! \xi_s$, $\bm{\eta}^{(0)} \!+\! \xi_h$ with $\xi_s \!\sim\! \mathcal{N}(0,I)$, $\xi_h \!\sim\! \mathcal{N}(0,0.64\,I)$ (the log-bandwidth variance $0.64 = 0.8^2$ was chosen empirically to cover roughly a factor-of-two bandwidth variation per restart).
\STATE Solve \eqref{eq:sensor_design} via L-BFGS-B $\to \theta_k^{(t)}$, $J^{(t)}$.
\ENDFOR
\STATE Select $\theta_k^\star \leftarrow \argmin_t J^{(t)}$.
\STATE Realized update: $w_{k,i}^+ \propto w_{k,i}^-\, L_k(x_{k,i}; \theta_k^\star)$.
\RETURN $\theta_k^\star$, $\{w_{k,i}^+\}$, fitting error $J^\star$, realized distance.
\end{algorithmic}
\end{algorithm}

\begin{remark}[Dimension-independence of the perturbation variance]
The log-bandwidth variance $0.64$ is a per-component quantity: each scalar $\eta_r$ is perturbed independently, so $e^{\eta_r + \xi}$ with $\xi \sim \mathcal{N}(0, 0.64)$ spans roughly $[h_r/2,\; 2h_r]$ regardless of the state-space dimension~$d$.
With diagonal or isotropic bandwidths, each coordinate of $\bm{\eta}$ receives the same perturbation, and the factor-of-two rationale applies entry by entry.
What does change with $d$ is the total parameter count, $(d+2)R$ for isotropic bandwidths or $(2d+1)R$ for diagonal, which reduces the probability that a random restart lands near a good basin.
The appropriate remedy is to increase the number of restarts~$T$, not to inflate the per-component variance.
\end{remark}

Since finite Gaussian mixtures are dense in the space of continuous likelihoods (in the supremum norm on compact sets), the realizability gap $\Delta_k(\theta_k) \to 0$ as the number of sensors $R$ increases, provided their positions, weights, and bandwidths are freely optimizable.

\paragraph{Integration into recursive state estimation.}
The sensor fitting procedure above determines the sensing architecture at a single time step.
In recursive state estimation the procedure is embedded in the standard predict-update loop: at each step $k$, the dynamical model propagates particles forward to produce the prior $\{x_{k,i}, w_{k,i}^-\}$; Layer~1 solves the maximum-entropy problem to determine the ideal likelihood $L_{k,i}^\star$; and Layer~2 fits the sensor parameters $\theta_k^\star$ to realize it.
The realizable filter update is then:
\begin{equation}
w_{k,i}^+ \propto w_{k,i}^-\, L_k(x_{k,i}; \theta_k^\star).
\label{eq:realizable_update}
\end{equation}

An important practical distinction governs how the two-layer pipeline is deployed.
In many platforms the sensor suite is fixed: the number, placement, and hardware characteristics of the sensors cannot be altered after installation.
For such systems the framework operates as an offline design tool.
The designer runs the two-layer pipeline over a representative ensemble of predicted priors and accuracy specifications, collects the optimal configurations $\{\theta_k^\star\}$, and selects a single static configuration (e.g., the one minimizing the worst-case realizability gap) that best serves the anticipated operating envelope.
Once installed, only the mixing weights $\bm{\alpha}_k$ and, if electronically adjustable, the measurement precision can be adapted online; the sensor positions $\bm{s}_k$ are frozen.

The framework realizes its full potential in dynamic sensing, where the location and precision of sensors can be reconfigured at each measurement interval.
Surveillance networks with mobile nodes, reconfigurable antenna arrays, and autonomous sensor platforms with controllable fields of view are natural applications.
In these settings, the fitted configuration $\theta_k^\star$ specifies where to place sensors, what precision each requires, and how to weight their contributions for the current interval.
As the state evolves, the optimal sensing architecture adapts: sensors reposition, bandwidths adjust, and mixing weights shift to track the changing likelihood shape dictated by the accuracy budget.
The operational cost term $\mathcal{J}(\theta_k, \theta_{k-1})$ in \eqref{eq:sensor_design} penalizes excessive reconfiguration, trading estimation accuracy against the physical cost of sensor repositioning.
This inverse approach naturally handles multimodal priors, asymmetric likelihoods, and support mismatch, because the target likelihood $L_k^\star$ inherits the geometry of the particle cloud at each step.
Algorithm~\ref{alg:sensor_design} summarizes the complete per-step procedure.

\begin{algorithm}[!ht]
\caption{Accuracy-Bounded Estimation with Sensor Design}
\label{alg:sensor_design}
\begin{algorithmic}[1]
\REQUIRE Prior particles $\{x_{k,i}, w_{k,i}^-\}_{i=1}^N$, desired posterior $\pi_k^\star$, budget $\varepsilon_k$, sensor family $\Theta$, previous positions $\bm{s}_{k-1}$.
\STATE \textbf{Layer 1 (information design):} Solve \eqref{eq:generic_particle_problem} $\to$ $w_k^{\mathrm{opt}}$.
\STATE \textbf{Target likelihood:} $L_{k,i}^\star \leftarrow w_{k,i}^{\mathrm{opt}} / w_{k,i}^-$.
\STATE \textbf{Layer 2 (sensor design):} Run Algorithm~\ref{alg:layer2} $\to$ $\theta_k^\star = (\bm{s}_k, \bm{\alpha}_k, \bm{h}_k)$.
\STATE \textbf{Realizable update:} $w_{k,i}^+ \propto w_{k,i}^-\, L_k(x_{k,i}; \theta_k^\star)$.
\STATE \textbf{Diagnostics:} Compute $D_{\KL}(w_k^{\mathrm{opt}} \| w_k^-)$ and $\Delta_k = \mathcal{D}(\pi_k^{+,\mathrm{real}}, \pi_k^\star) - \varepsilon_k$.
\STATE \textbf{Resample} if $\ESS(w_k^+) < N_{\mathrm{thresh}}$.
\RETURN $\{x_{k,i}, w_{k,i}^+\}$, sensor config $\theta_k^\star$, diagnostics $(D_{\KL}, \Delta_k)$.
\end{algorithmic}
\end{algorithm}

\subsection{Design Diagnostics}\label{sec:diagnostics}

Two quantities serve as design criteria at each step:
\begin{enumerate}
\item \emph{KL injection:} $D_{\KL}(w_k^{\mathrm{opt}} \| w_k^-)$ measures the information cost of meeting the accuracy budget.
A large value indicates that the budget $\varepsilon_k$ is tight relative to the prior-target gap; the designer should either relax $\varepsilon_k$ or improve the prior through better dynamics modeling.
When the prior is already close to the target, the injection is small and the constraint may be inactive.
\item \emph{Realizability gap:} $\Delta_k = \mathcal{D}(\pi_k^{+,\mathrm{real}}, \pi_k^\star) - \varepsilon_k$ measures how well the physical sensor family approximates the ideal likelihood.
If $\Delta_k > 0$, the sensor family is too restrictive: either add sensors (increase $R$), allow broader placement regions, or relax the budget.
If $\Delta_k \le 0$, the accuracy specification is achievable with the current sensor suite.
\end{enumerate}
These two quantities separate the information-theoretic and physical failure modes: the budget may demand too much information (high KL), or the sensor family may lack the degrees of freedom to deliver it (positive $\Delta_k$).

% ============================================================
\section{Numerical Experiments}\label{sec:experiments}

We compare four accuracy metrics, Wasserstein ($W_2$), MMD, moment constraints, and $\chi^2$-divergence (an $f$-divergence), on two scenarios that span the range from straightforward to structurally challenging.
Each metric is placed in the same framework: minimize $D_{\KL}(\pi \| \pi_k^-)$ subject to the respective accuracy constraint, with tolerance parameters chosen so that each metric can visibly depart from the prior.
We use $N = 2000$ prior particles and $M = 500$ target particles.

\begin{figure}[t]
\centering
\includegraphics[width=\columnwidth]{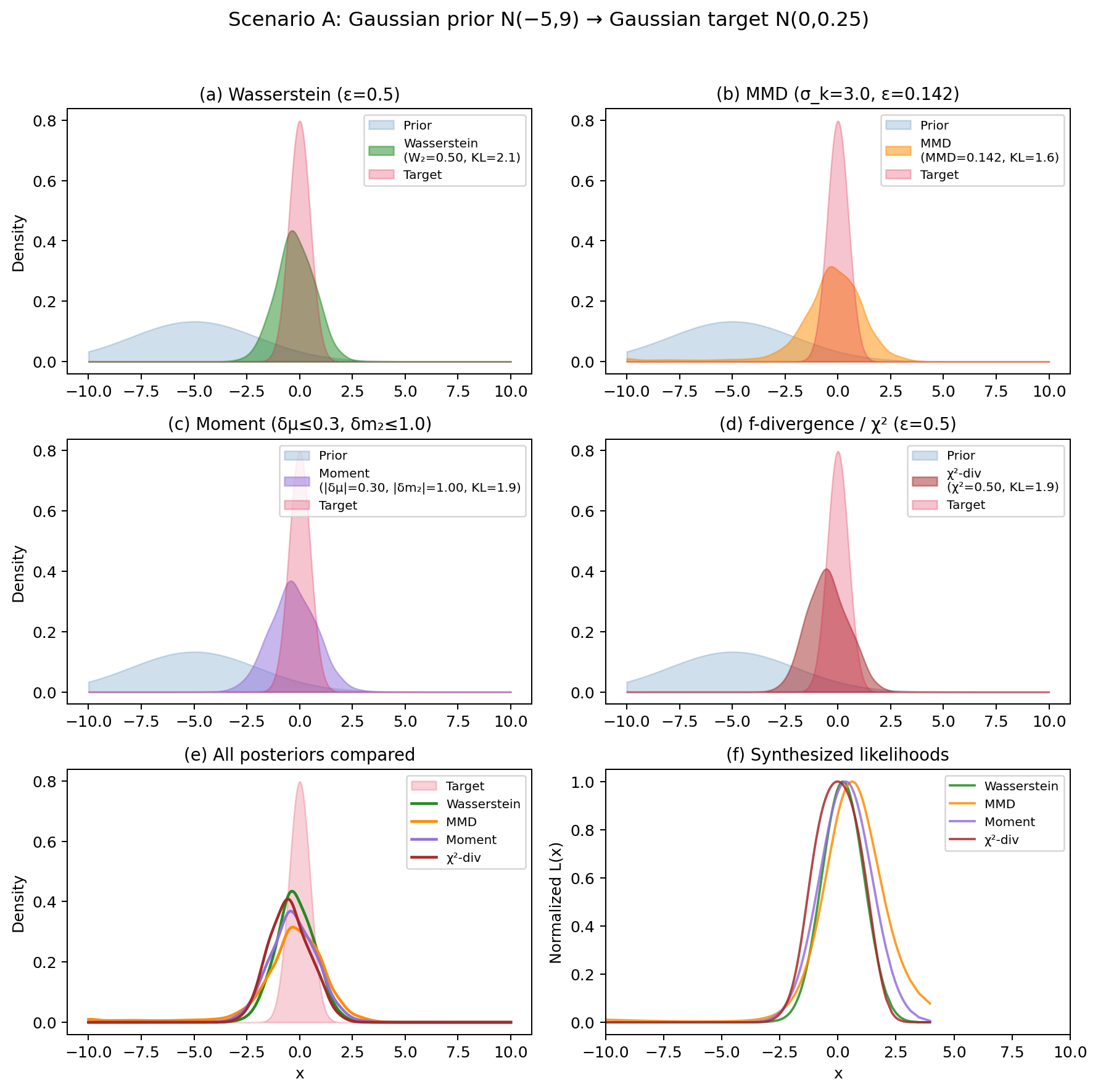}
\caption{Scenario~A (Gaussian to Gaussian): four accuracy metrics applied to unimodal prior and target.  Panels (a)--(d) show individual posteriors; panel~(e) overlays all posteriors; panel~(f) shows synthesized likelihoods.  Metric choice has limited effect when both distributions are unimodal.}
\label{fig:exp3a}
\end{figure}

\paragraph{Scenario~A: Gaussian to Gaussian.}
The prior is $\pi^- = \mathcal{N}(-5, 9)$ and the target is $\pi^\star = \mathcal{N}(0, 0.25)$.
Fig.~\ref{fig:exp3a} shows the results.
All four posteriors shift toward the target with similar shape, differing primarily in variance. Wasserstein and $\chi^2$-divergence yield the tightest posteriors at highest KL cost ($\approx 2$); moment constraints produce a broader posterior at lower KL ($1.88$); MMD retains the most spread, consistent with kernel smoothing.
The synthesized likelihoods are nearly identical, indicating limited metric dependence for unimodal problems.

\begin{figure}[t]
\centering
\includegraphics[width=\columnwidth]{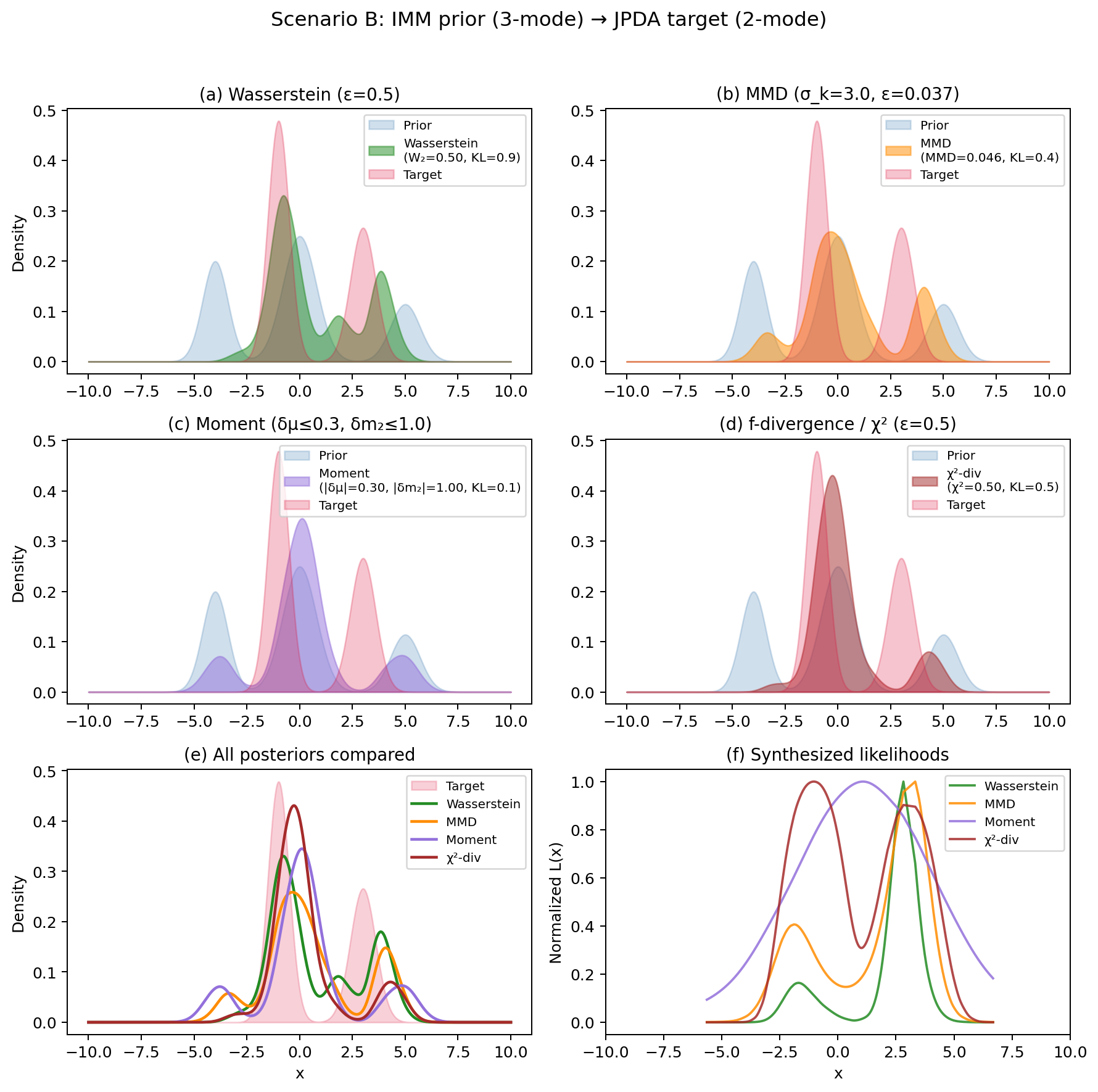}
\caption{Scenario~B (multimodal to multimodal): The four metrics produce qualitatively different posteriors and likelihoods.  Wasserstein concentrates mass near target modes; moment constraints barely move the prior; MMD and $\chi^2$-divergence produce intermediate but distinct shapes.}
\label{fig:exp3b}
\end{figure}

\paragraph{Scenario~B: Multimodal to multimodal.}
The prior is a three-component Gaussian mixture with weights $(0.5, 0.3, 0.2)$, means $(0, -4, 5)$, and standard deviations $(0.8, 0.6, 0.7)$.
The target is a two-component mixture with weights $(0.6, 0.4)$, means $(-1, 3)$, and standard deviations $(0.5, 0.6)$.
Mode locations do not coincide, forcing the metric to reconcile structurally unrelated mixtures.
This scenario is motivated by IMM tracking \cite{Blom1988,BarShalom2001} and multi-hypothesis data association \cite{Fortmann1983,BarShalom2001}, where multimodal densities arise from motion hypotheses or measurement ambiguity.
Fig.~\ref{fig:exp3b} shows the results.
Now the four metrics produce qualitatively different posteriors:
\begin{itemize}
\item \emph{Wasserstein} ($W_2 = 0.50$, $D_{\KL} = 0.95$) transports mass geometrically toward the target modes, achieving the best distributional accuracy but injecting the most information.
\item \emph{MMD} ($\MMD = 0.046$, $D_{\KL} = 0.39$) smooths through the kernel and produces a broad posterior that roughly covers the target support without resolving its bimodal structure.
\item \emph{Moment constraints} ($|\delta\mu| = 0.30$, $|\delta m_2| = 1.0$, $D_{\KL} = 0.14$) match the target mean and second moment while barely departing from the prior, yielding the most conservative update but the worst geometric fit.
\item $\chi^2$\emph{-divergence} ($\chi^2 = 0.50$, $D_{\KL} = 0.48$) penalizes pointwise density-ratio deviations, producing the tightest posterior variance ($3.1$, undershooting the target's $4.3$) with a shape distinct from all other metrics.
\end{itemize}
The synthesized likelihoods (panel~f) are strikingly different, confirming that in structurally challenging problems the choice of accuracy metric determines not just how much information is injected, but where that information concentrates.
The accuracy thresholds for each metric ($\varepsilon_W$, $\varepsilon_{\MMD}$, $\varepsilon_\mu$, $\varepsilon_{m_2}$, $\varepsilon_{\chi^2}$) were chosen ad hoc for demonstration purposes.

\begin{figure}[t]
\centering
\includegraphics[width=\columnwidth]{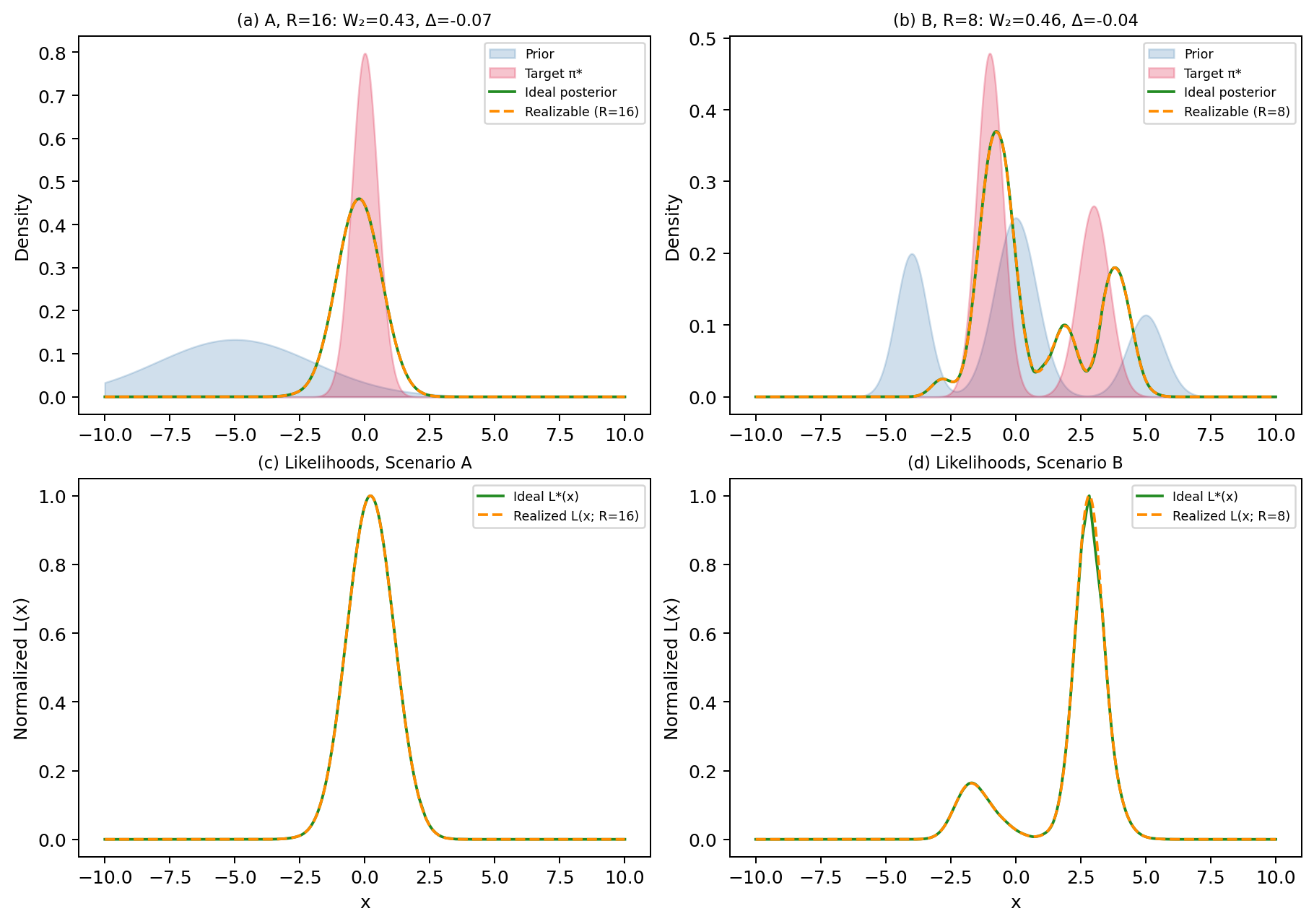}
\caption{Sensor realizability.  Top: prior (blue), target (red), ideal posterior (green), realized posterior (orange dashed).  Bottom: ideal likelihood $L^\star$ (green) versus realized $L(\cdot\,; \theta^\star)$ (orange dashed).  Each column shows best~$R$ for that scenario.  Realized likelihood closely approximates $L^\star$, confirming the two-layer architecture translates budgets into achievable configurations.}
\label{fig:sensor_design}
\end{figure}

\paragraph{Sensor realizability.}
To demonstrate sensor realizability, we apply the full two-layer pipeline (Algorithms~\ref{alg:sensor_design} and~\ref{alg:layer2}) to both scenarios using the Wasserstein metric $W_2$ with budget $\varepsilon = 0.5$, sweeping the number of sensors from $R = 1$ to $16$; the same procedure applies to any of the other accuracy metrics.
Layer~1 produces ideal weights $w^{\mathrm{opt}}$ and implied likelihood $L_i^\star = w_i^{\mathrm{opt}} / w_i^-$; Layer~2 fits the parametric sensor likelihood \eqref{eq:sensor_mixture} via Algorithm~\ref{alg:layer2} with $T = 8$ restarts.
Fig.~\ref{fig:sensor_design} shows results for the best~$R$ in each scenario.

In Scenario~A, the ideal likelihood has a single Gaussian-shaped peak; even $R = 1$ achieves near-zero fitting error and realized $W_2 = 0.43$ (below the ideal $0.50$).
The realized posterior is visually indistinguishable from the ideal.
Adding sensors produces no further improvement.

Scenario~B is more demanding.
The ideal likelihood is bimodal.
A single sensor cannot capture this: fitting error $1.22$, $W_2 = 1.64$.
With $R = 2$, fitting error drops to $0.03$ and realized $W_2 = 0.49$ matches the ideal.
For $R \ge 4$ the fit saturates.
These results confirm that realizability gap $\Delta_k$ decreases with $R$ and vanishes once the sensor family has sufficient degrees of freedom.

% ============================================================
\section{Discussion and Extensions}\label{sec:discussion}

\paragraph{Online adaptation of $\varepsilon_k$.}
The accuracy budget can be adapted online: tighten $\varepsilon_k$ when the ESS is high and KL injection is small (the filter is operating conservatively), and relax it when the ESS drops sharply to prevent particle depletion.

\paragraph{Scalability.}
The main computational bottleneck is the Wasserstein constraint evaluation via Sinkhorn, scaling as $\mathcal{O}(NM)$ per iteration.
For large particle counts, stabilized sparse scaling \cite{Schmitzer2019} or multiscale Sinkhorn methods \cite{Peyre2019} could reduce this cost.
The MMD and moment-constraint alternatives offer better scaling when transport-level accuracy is not required.

\paragraph{Relationship to optimal transport.}
The Sinkhorn solver used for constraint evaluation is differentiable \cite{Cuturi2013}, which opens a path toward end-to-end gradient-based sensor optimization through the transport layer.

\paragraph{Higher dimensions.}
All experiments in this paper are one-dimensional.
The framework itself is dimension-agnostic: the particle-level programs \eqref{eq:generic_particle_problem}--\eqref{eq:wasserstein_program} operate on weighted point clouds in $\mathbb{R}^d$ without modification.
However, practical challenges grow with $d$: the particle count $N$ needed for adequate coverage scales exponentially, the Sinkhorn cost matrix becomes $\mathcal{O}(NMd)$ to fill, and the sensor likelihood \eqref{eq:sensor_mixture} requires multivariate bandwidths (full or diagonal covariance per component) whose parameterization adds $\mathcal{O}(Rd^2)$ or $\mathcal{O}(Rd)$ variables.
For moderate dimensions ($d \le 5$), these costs remain manageable; for high-dimensional problems, projection-based or sliced-Wasserstein approximations may be necessary.

% ============================================================
\section{Conclusion}\label{sec:conclusion}

We have presented a maximum-entropy framework for synthesizing likelihoods in particle filtering when forward measurement models are unavailable.
The formulation, KL minimization subject to distributional accuracy constraints, decouples accuracy specification from information injection.
Instantiations for Wasserstein, MMD, $f$-divergences, moment constraints, and hybrid metrics \eqref{eq:multi_constraint} yield convex (or convex-relaxed) programs, with a closed-form exponential-tilt solution for the symmetric case.
The desired posterior is a performance specification, not an oracle; in the simplest case, the budget is a direct RMS error bound.
A two-layer sensor design architecture uses the synthesized likelihood as the interface between estimation requirements and physical sensor configuration.
Experiments comparing Wasserstein, MMD, moment constraints, and $\chi^2$-divergence confirm reliable constraint enforcement and show that metric choice has limited impact for unimodal problems but produces qualitatively different posteriors and likelihoods when the prior and target are multimodal.
A sensor realizability experiment demonstrates that the parametric likelihood family can closely approximate the ideal likelihood shape, with the realizability gap vanishing as the number of sensors increases.

Future work includes sequential tracking validation over multiple time steps, formal approximation rates for the realizability gap as a function of sensor count, comparison with classical sensor design methods (D-optimal, Fisher information, mutual information), online budget adaptation, and scaling to higher dimensions.

% ============================================================

\end{document}